\documentclass[twocolumn]{article}

\usepackage{graphicx} 
\usepackage{amsmath} 
\usepackage{authblk}

\usepackage[usenames, dvipsnames]{xcolor}
\def \revision#1{{\color{black}#1}}

\usepackage[margin=2cm, top=2cm, bottom=2cm]{geometry}

\title{\textbf{Discovery of propagating trains of oscillons over Faraday waves in a 1D experiment}}

\author[1,2,4]{S. Kucher}
\author[2,4]{J. E. Wesfreid}
\author[1,3,4]{P. J. Cobelli}
\affil[1]{\small\textit{Universidad de Buenos Aires, Facultad de Ciencias Exactas y Naturales, Departamento de F\'isica, Ciudad Universitaria, 1428 Buenos Aires, Argentina}}
\affil[2]{\textit{Laboratoire de Physique et M\'ecanique des Milieux H\'et\'erog\`enes, UMR CNRS 7636, ESPCI-Paris, Universit\'e PSL, Sorbonne Universit\'e, Universit\'e Paris Cit\'e, 75005 Paris, France}}
\affil[3]{\textit{CONICET - Universidad de Buenos Aires, Instituto de F\'isica Interdisciplinaria y Aplicada (INFINA), Ciudad Universitaria, 1428 Buenos Aires, Argentina}}
\affil[4]{\textit{IRL2027 - Institut Franco-Argentin de Dynamique des Fluides pour l'Environnement (IFADyFE)}}

\date{}

\begin{document}


\twocolumn[{
\begin{@twocolumnfalse}
  \maketitle
  
  \begin{abstract}
    We report the discovery of highly localized structures traveling over a one-dimensional pattern of Faraday waves in a vertically-vibrated fluid layer confined in a thin annular cell. These propagating structures emerge spontaneously beyond a threshold and coexist with the underlying pattern. They move at constant speed, in trains of sharp peaks that co- and counter-propagate along the cell, with velocities largely exceeding the Faraday waves drift. Our results raise the question whether propagating localized structures are also observable in other parametrically driven systems in physics.
  \end{abstract}
  
  \vspace{1em} 
\end{@twocolumnfalse}
}]

\section{Introduction}

The emergence of patterns in dissipative systems driven far from equilibrium
is a universal phenomenon that bridges the realms of the most diverse scientific disciplines. From morphogenesis in developmental biology \cite{Turing1952,Kondo2010}, cloud formation in meteorology \cite{Monroy2021,Rosemeier2021}, reaction-diffusion systems in chemistry \cite{Lindgren2024}, to neural dynamics in brain function \cite{Hutt2014}, highway traffic \cite{Eddie2008}, directed networks \cite{Asllani2014}, nonlinear optics \cite{Arecchi1999}, condensed matter physics \cite{Mouritsen1990}, fluid dynamics \cite{WesfreidBook1984,wesfreid1987propagation,cross2009pattern}, cosmology \cite{Liddle_Lyth_2000}, and quantum systems \cite{Zhang2019}, pattern formation plays a central role in shaping our understanding of complex systems \cite{hoyle2006pattern,pismen2006patterns,walgraef2012spatio}.%

Patterns often span the entire system, but they can also appear as spatially localized structures embedded in a homogeneous, fluctuating or even noisy background, as well as in periodically driven systems \cite{Tlidi2014,Knobloch2015}. A paradigmatic example of the former is Faraday waves, in which a shallow layer of fluid is subject to vertical oscillations, the acceleration periodically modulating the effect of gravity. Beyond a driving threshold, a subharmonic surface wave instability develops which ultimately leads to the emergence of a spatial pattern in the form of crispations of the free-surface \cite{Faraday,Miles1993}. Almost two centuries
after its discovery, this pattern-forming parametric wave instability continues to offer novel insights and valuable connections between
various domains of physics. For instance, Faraday waves are fundamental to the motion of walking droplets \cite{Couder2005,bush2020hydrodynamic, thomson2020}, and have been observed in nonlinear optics \cite{Perego2018}, in strongly interacting superfluids \cite{HernandezRajkov2021}, and in Bose-Einstein condensates of ultracold quantum gases \cite{Engels2007,Nguyen2019}. Their study has also led to technological applications such as particulate film patterning \cite{Wright2003}, biomedical microfabrication and assembly \cite{Guex2021,Nian2023,Tognato2023} and drug delivery \cite{Tsai2017}.

\begin{figure}[!t]
    \centering
    \includegraphics[width=\linewidth,]{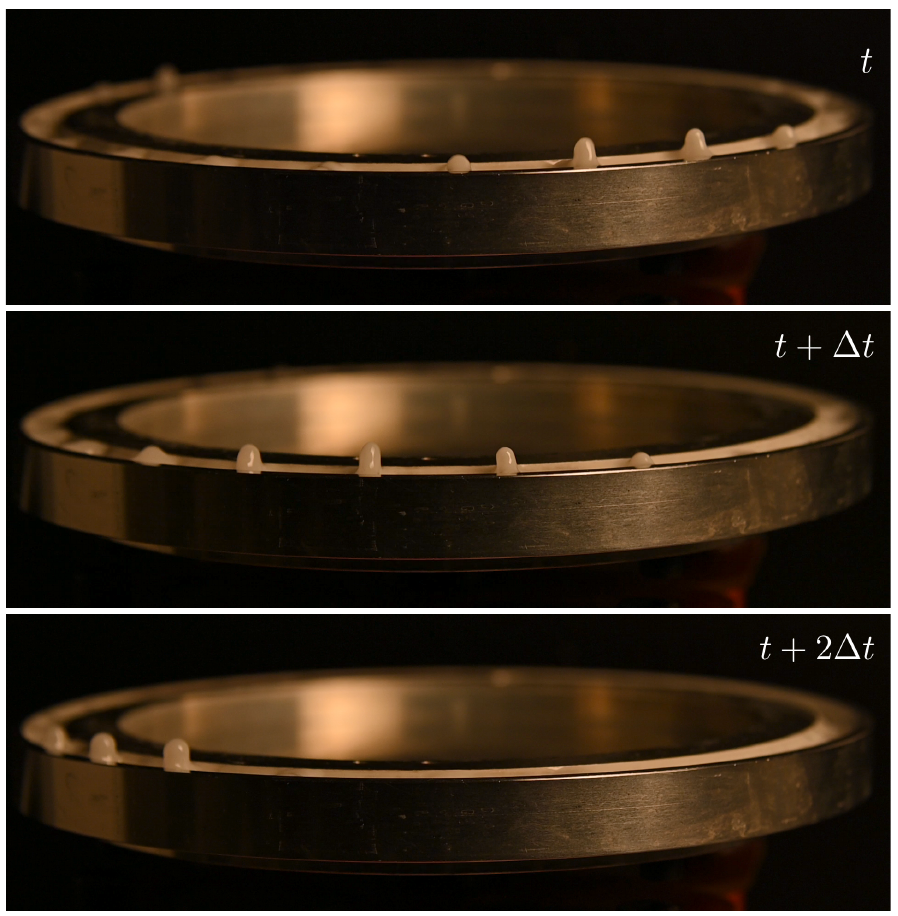}
    \caption{Time evolution (top to bottom) of a train of localized structures propagating clockwise over a spatially-periodic background of Faraday waves, as observed in our experimental setup. Snapshots, separated by $\Delta t = 0.75$~s, correspond to an experiment with $\epsilon = 1.10$. A movie is available in the Supplementary Material.}
    \label{fig:snapshots}
\end{figure}

It has been shown that parametrically driven surface waves admit the formation of localized structures, known as \textit{oscillons} \cite{Ibrahim2021}. They have been first observed in granular matter \cite{Umbanhowar1996, Wu1984} and later in vertically excited fluid layers. Generally speaking, these structures can be of two different types depending on the dynamics that lead to their formation. On one hand, a solitary state can appear on a flat surface, presenting a few isolated peaks  \cite{Fineberg1998, Lioubashevski1999} and as rogue wave \cite{dysthe2008oceanic, Xia2012}.  This type of structure is produced by a subcritical hysteretical bifurcation \cite{Pradenas2017}. They have been reported both in two dimensional experiments as well as in one dimensional ones \cite{Rajchenbach2011, Li2014}. Recently, localized structures were observed with a spatially inhomogeneous Faraday forcing  \cite{Urra2019,marin2023drifting}. On the other hand, localized structures have been observed on top of an underlying bidimensional pattern \cite{Cabeza2003, Xia2012}.  This type of oscillons has been observed on top of a bidimensional Faraday pattern, with localized peaks moving chaotically while preserving their shape \cite{Xia2012}. 

In this Letter we report the discovery of trains of propagating localized structures over Faraday waves in a vertically-vibrated fluid layer confined within a thin annular cell.  Figure \ref{fig:snapshots} shows an example of such structures and their dynamics. These superstructures take the form of several peaks or \textit{oscillons} that propagate on top of the underlying spatially-periodic background made of by Faraday waves, and are characterized by heights that significantly exceed that of the background. This amplitude difference between the localized structures and the background waves can be seen not only with the naked eye, but more prominently after data analysis (Figures \ref{fig:lines} and \ref{fig:xt_osc}).
We observe a remarkable spatial extension and organization in addition to its propagating trend.

\section{Experimental setup}

\begin{figure}[!t]
    \centering
    \includegraphics[width=0.8\linewidth,]{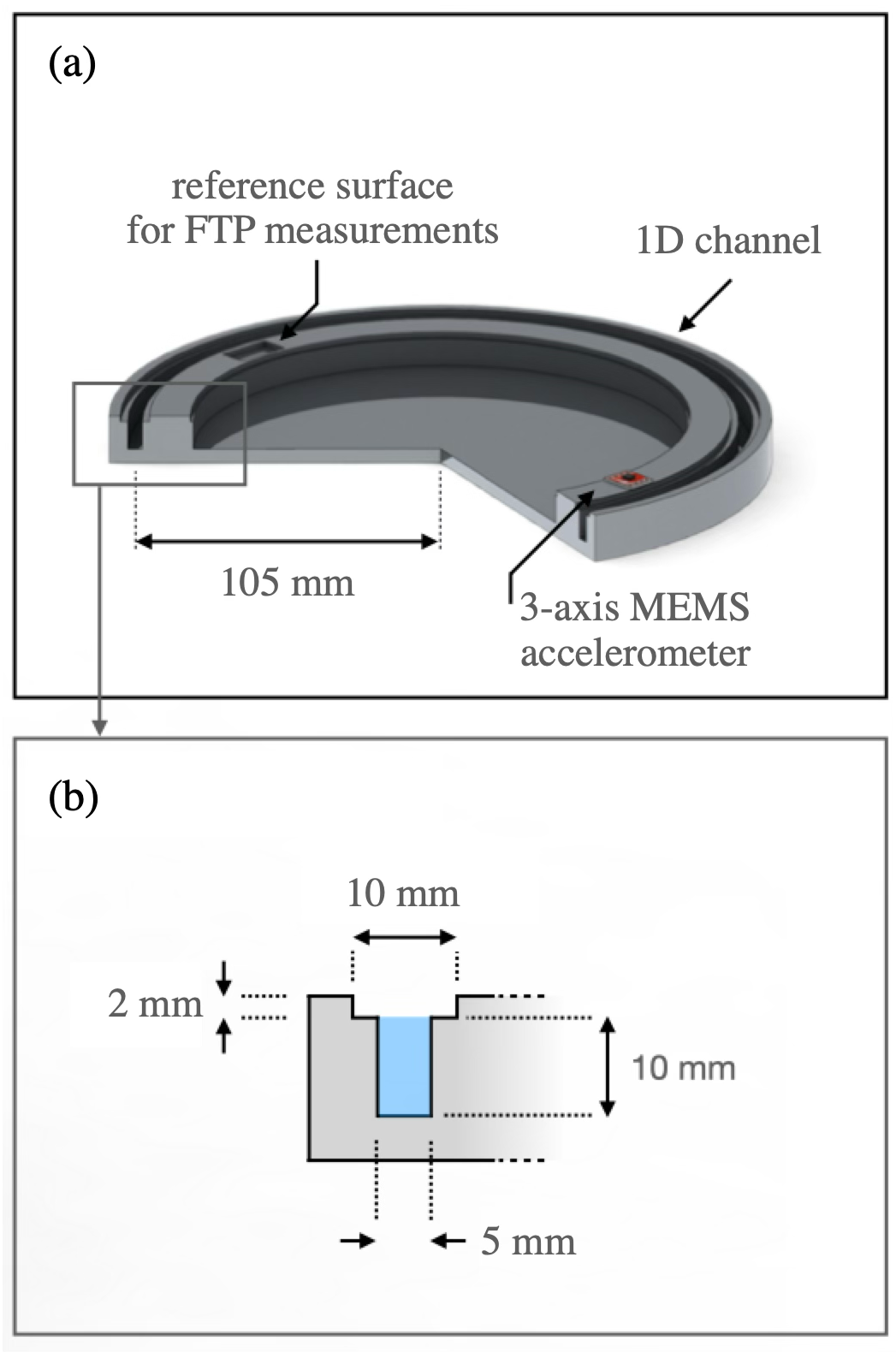}
    \caption{Schematic diagram of the experimental setup, showing (a) the cell details and (b) the 1D channel geometry.}
    \label{fig:exp_setup}
\end{figure}

Our experimental system is represented schematically in Fig.~\ref{fig:exp_setup}. It comprises a 10~mm deep annular channel of square section (210~mm mean diameter; $5$~mm width) filled with distilled water. 

The fluid container is mounted on a Br\"uel \& Kjaer LDS V406 electromagnetic shaker driven by a LDS LPA100 signal amplifier connected to an Agilent 33220A waveform generator (not shown). The cell itself is comprised of two separate regions for holding liquids: the inner disk (not used in this study) and the outer annulus or 1D channel of rectangular cross-section in which our experiments were performed. In Fig.~\ref{fig:exp_setup}(b), the portion of the channel that the fluid occupies is marked by a light blue rectangle, corresponding to a cross-section of area $(10 \times 5)$~mm$^2$. The channel is filled to its brim in order to pin the free-surface at the edge of the channel sidewalls, preventing problems arising from the curvature of the interface at the boundaries \cite{Douady1990}. Since water is sensitive to ambient contamination \cite{Henderson1998EffectsOS, MARTÍN_VEGA_2006}, we renewed the test fluid completely before each set of measurements to ensure a clean, uncontaminated surface.

The dynamics of the free-surface relative to the moving container is measured by Fourier Transform Profilometry (herein, FTP) \cite{Cobelli2009, Maurel2009}. 
In the framework of this technique, a sinusoidal fringe pattern of known characteristics is projected onto the surface whose shape is interrogated, and its image is registered by a camera. Perturbations in the air-liquid interface introduce local frequency modulations in the observed pattern with respect to the projected one. Their phase difference leads to a phase map from which the local free surface height can be reconstructed. 
In order to be able to project a pattern onto the liquid's surface, we use a dilute
suspension of TiO$_2$ particles (Kronos 1001) in distilled water (3~g/l) as
the working liquid. This allows us to dye the water white without significantly
modifying its hydrodynamical properties \cite{Przadka2012}. 
We have also checked that our results are also observable with the use of
distilled water without TiO$_2$ particles.

A 3LCD FullHD Epson TW1000 videoprojector placed over the cell is employed to
project a pattern of known characteristics onto the free surface. The fluid
surface is imaged from above using a ProcImage500-Eagle CMOS camera ($1280
\times 1024$~px$^2$ resolution, 250~fps).  The FTP system composed of the 
videoprojector--camera pair is arranged in parallel-optical-axes geometry. 
For the purpose of measuring the deformation of the free-surface {\it relative to the moving
container}, the cell was designed to provide a flat surface precisely machined
at the same height as that of the free surface at rest, to serve as an
instantaneous height reference with respect to which the instantaneous free
surface deformation is measured. 

A 3-axis ADXL345 MEMS accelerometer ($\pm 4 g$, 10 bit resolution, 3200 Hz acquisition frequency) is attached to the cell and controlled externally by an Arduino DUE board. This device serves the two-fold
purpose of recording the instantaneous vertical acceleration as well as
monitoring the horizontal motion to ensure no sloshing motion is present. In
the experiments, the in-plane acceleration was checked to be Gaussian noise with zero mean; the horizontal acceleration was found to be less than 1\% of the vertical one.

The channel is mounted on an electromagnetic shaker that vibrates vertically with a fixed harmonic excitation of frequency $f = 20$~Hz. Only the amplitude of the forcing is varied to set the vessel maximum acceleration in each experiment. Similar results were observed for $f = 15$~Hz (see Supplementary Material).

As the channel width is smaller than the wavelength corresponding to the fundamental transverse (radial) mode for gravity-capillary waves, it can be regarded as a one-dimensional system in which only the azimuthal coordinate is relevant. We calculate the average height in the annular region by integrating over the radial direction, i.e., $h(\theta, t) = \langle h(r,\theta,t) \rangle_{r}$

In the following, our observations are discussed in terms of the degree of supercriticality above the threshold acceleration, denoted by $\epsilon = a / a_{c} -1$. Here, $a$ represents the driving acceleration, while $a_{c} = (0.62 \pm 0.01)~g$ corresponds to the onset of the Faraday instability in our experimental setup; $g$ being the local gravity acceleration. 
We varied $\epsilon$ in the range $[0.05, 1.33]$, in 14 steps of approximately $0.10$.

\begin{figure}[!t]
    \centering 
    \includegraphics[width=\linewidth]{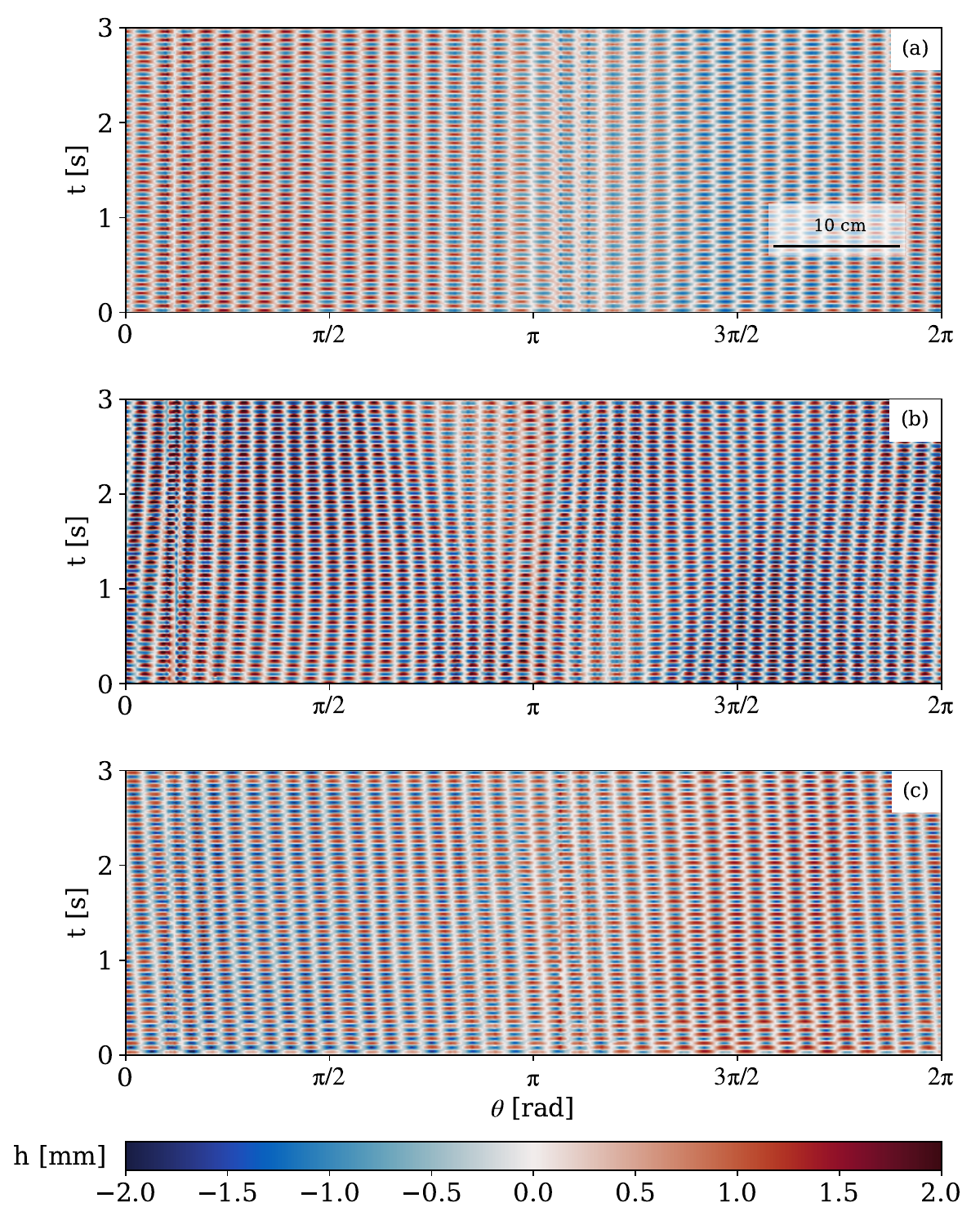}
    \caption{
    Spatiotemporal diagrams of the free surface elevation $h(\theta,t)$ showing the different observed behaviors: (a) Faraday waves, (b) vacillating-breathing states, and (b) slow drift of the Faraday pattern.}
    \label{fig:xt_faraday}
\end{figure}

\section{Behavior of Faraday patterns}

In the following, we briefly describe the different behaviors we observed for the system. For $0 < \epsilon < 0.65$ only the Faraday waves are observed,
their amplitude increasing progressively with the forcing. For values of $\epsilon$ above $0.65$, we start to observe a drift of the Faraday pattern. For $\epsilon \geq 0.73$, we also observe vacillations--breathing, i.e., a local modulation of the wavelength. For $\epsilon \geq 0.93$, pulses appear, and may coexist with all the above. Moreover, they become more stable as the forcing grows.

Figure~\ref{fig:xt_faraday} shows examples of spatiotemporal diagrams $h(\theta, t)$ for three different chosen values of the forcing.
The top panel, corresponding to $\epsilon = 0.65$, clearly shows the presence of Faraday waves, which are evident from the periodicity in both space and time revealed in the checkerboard-like structure of the spatiotemporal diagram. As expected, and previously observed \cite{Kumar1994, Benjamin2006}, they exhibit a subharmonic response of frequency $f/2$; their wavelength being $\lambda = (34.7 \pm 1.7)\ \textrm{mm}$.

As the driving amplitude is increased beyond this value, the standing wave pattern presents an expansion-compression mode of the periodic structure; i.e., a wavelength modulation both in space and time. This phase modulation of the periodic pattern was previously observed \cite{Douady1989} as a secondary instability of parametrically generated surface waves, located close to the drift bifurcation in the experimental parameter space. This is shown in Fig.~\ref{fig:xt_faraday}(b) ($\epsilon = 0.74$) in the form of compressions and expansions of the Faraday pattern, corresponding to modulations of its local  wavelength.

For $\epsilon = 0.93$ (Fig.~\ref{fig:xt_faraday}(c)), the system has transitioned a secondary instability in the form of a drift of the Faraday pattern, manifested as a slight shear of the checkerboard motif. The entire structure moves slowly at constant speed, in agreement with what was reported in \cite{Douady1989}. This drift showed no preference in sense, as it alternated randomly between clockwise and counter-clockwise in repeated experiments. The drift velocity is $c_{\text{d}} = (4.7 \pm 0.2) \ \textrm{mm/s}$; i.e., taking approximately $140$~s for the pattern to complete one revolution around the channel.

\section{Observation of localized propagating structures}

At later stages (and also for higher levels of forcing), we observe the spontaneous emergence of highly localized structures in the form of sharp peaks of height much larger than the amplitude of the Faraday waves they coexist with. These oscillons propagate at a constant speed along the annulus while oscillating at half the driving frequency, and appear in groups (trains).

\begin{figure}[!t]
    \includegraphics[width=\linewidth]{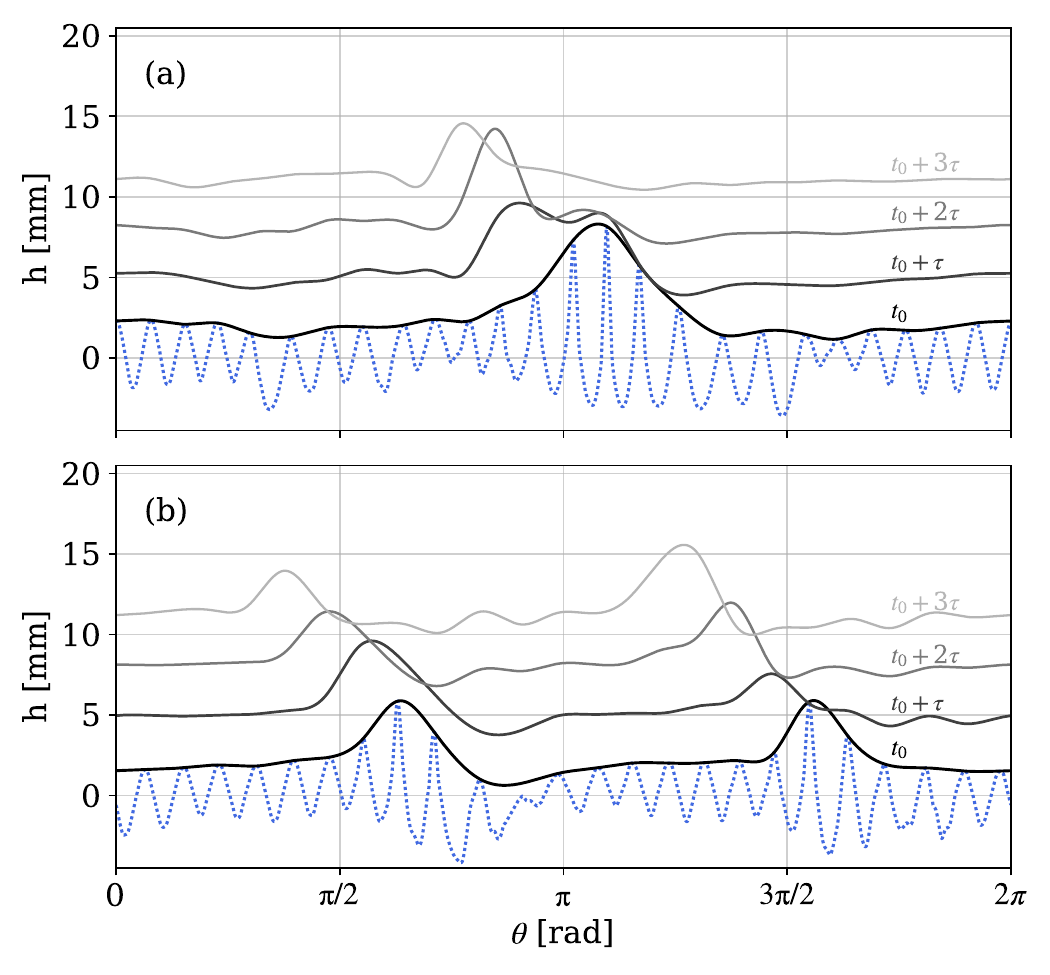}
    \caption{Height profiles of the free-surface as measured by FTP, showing the propagation of super-peaks over the periodic Faraday background. Panel (a) shows a case with only one train, whereas two trains are displayed in panel (b). Both panels correspond to different experimental realizations with the same value of $\epsilon = 0.93$.}
\label{fig:lines}
\end{figure}

Instantaneous height profiles for two different experimental realizations at the same forcing level ($\epsilon = 0.93$) are displayed in dotted (blue) lines on Figure~\ref{fig:lines}. For these, the zero-height level corresponds to the position of the free-surface at rest.
The shape of the free-surface profiles (dotted lines in the figure) reveals the presence of nonlinearity in the system. This is evidenced by the top-bottom asymmetry resulting from the influence of higher harmonics. Consequently, the wave pattern exhibits sharp crests and flat troughs, far from sinusoidal Faraday waves. Due to the top-bottom asymmetry of the dotted profiles, we  introduce an upper envelope to quantify the presence of local amplitude modulation.

In both panels, the continuous black lines represent the upper envelopes of each instantaneous height profiles, calculated as $2\pi$-periodic cubic spline interpolants connecting adjacent local maxima. Envelopes for later times, separated by 1$\tau$, 2$\tau$, and 3$\tau$ (with $\tau = 2/f = 0.10$ s) are also shown in the panels, shifted vertically for the sake of clarity.
The upper and lower panels show the occurrence of one and two co-propagating trains of oscillons, respectively. In the former, the oscillons attain a maximum height of 8.3 mm; i.e., more than three times larger than the amplitude of the periodic Faraday background. The presence of a second train in the lower panel leads to a natural decrease in peak height compared to panel (a), as the same injected energy is now distributed between both trains (and the Faraday waves). Nevertheless, the peaks in the lower panel still remain significantly higher than the amplitude of the local Faraday pattern. 

Another aspect of the localized structures that becomes evident in this representation is the left-right asymmetry around each peak of the upper envelopes. This distinctive spatial variation in the characteristics of the localized structures deserves further investigation, as it may potentially offer valuable insights into their underlying dynamics and formation processes.

\begin{figure}[!t] 
    \includegraphics[width=0.9\linewidth]{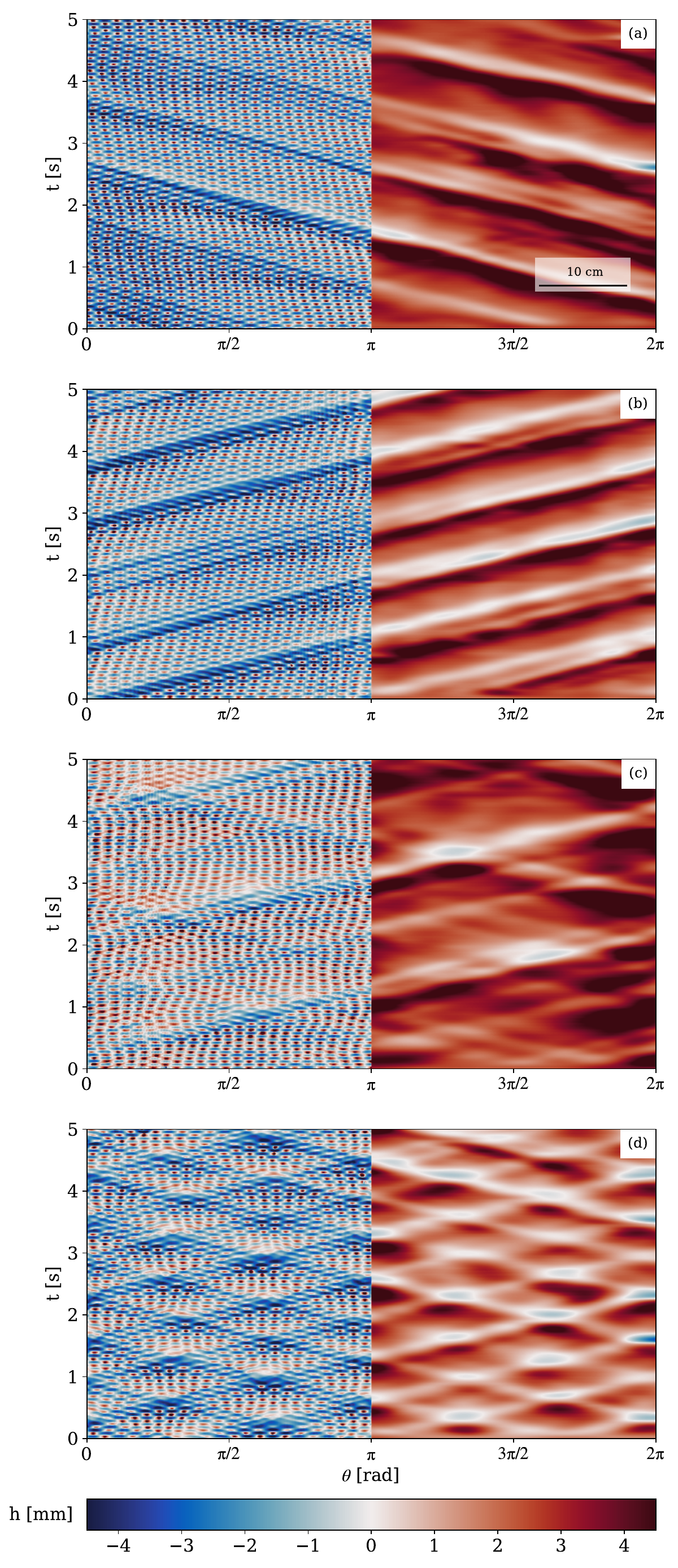}
    \caption{Spatiotemporal diagrams of the free-surface height (left halves) and corresponding upper envelopes (right halves) showing the typical behaviors observed: (a) counterclockwise propagation, (b) clockwise propagation, (c) propagation in both senses  and (d) simultaneous propagation of three trains in opposite directions. Panels (a), (b) and (c) correspond to a forcing level  of $\epsilon=0.93$, while (d) corresponds to $\epsilon=1.32.$ }
\label{fig:xt_osc}
\end{figure}

In terms of the envelopes, Figure~\ref{fig:lines} illustrates the occurrence of localized structures propagating over a non-null homogeneous background of Faraday waves. We have observed the emergence of these localized propagative structures in a number of different configurations, including, but not limited to, individual and simultaneous counter- and co-propagation of two or more trains of oscillons.

Figure~\ref{fig:xt_osc} presents a summary of the behaviors observed throughout our study. Each panel shows a composite spatiotemporal diagram made of the free-surface dynamics (left half; $0 \leq \theta \leq \pi$) and of its corresponding upper envelope filtered in frequency to remove the Faraday waves (right half; $\pi < \theta \leq 2\pi$). As a whole, these four panels depict the propagation of these structures in two directions (clockwise and counter-clockwise) on the cell, both separately and simultaneously. All panels depict the system behavior within a 5-second time window, corresponding to 100 cycles of the driving.

Figure~\ref{fig:xt_osc}(a) presents a composite spatiotemporal diagram for a measurement at a forcing level of $\epsilon~=~0.93$, depicting two trains, located on opposite sides of the channel, and traveling counterclockwise along it. As these narrow structures are surrounded by wide troughs, their footprints in the spatiotemporal diagram (left half) are noticeable as blue scars that cut through the underlying Faraday pattern.

In this measurement, two oscillons propagate in the same direction at a constant speed, with a mean velocity of $c~=~(307 \pm 6)$~{mm/s}. The left half of the diagram also displays the slow drifting of the underlying Faraday pattern, which coexists with the passage of the oscillons. Notably, their relative velocity is significantly larger than the drift velocity, as evidenced by the structure completing an entire revolution around the vessel in 2.15 seconds (movies are available in the Supplementary Material). 
For the same level of forcing, we also observed the propagation of such structures in the opposite direction, as shown in Fig.~\ref{fig:xt_osc}(b). It is interesting to note that the \revision{group} velocity for gravity-capillary waves at $f$ = 10 Hz (subharmonic) is \revision{$c_{\text{group}} = \partial \omega /\partial k = 205$ mm/s}, comparable to the speed of the oscillons and two orders of magnitude larger than the Faraday waves drift.

In the two cases discussed up to this point (panels a and b), the emergence of unidirectional localized propagating structures may seem to involve a spontaneous symmetry breaking in which the system passes from one symmetric state, i.e., that of the Faraday waves, to another in which one direction is preferred. This symmetry breaking occurs in both directions for different realizations at the same experimental conditions; indicating that the sense of propagation for such structures is chosen randomly by the system. An additional observation concerning this point can be drawn from comparing panels (a) and (b). Within each diagram, the underlying drift of the Faraday pattern and the direction of propagation of the localized structures coincide, pointing to the fact that the system is developing these structures from an already asymmetric state: that in which the preexisting Faraday drift conditions the direction of propagation. 

At the same level of forcing ($\epsilon = 0.93$), we also observed simultaneous propagation of these localized structures in opposite directions. An example of such behavior is shown in Fig.~\ref{fig:xt_osc}(c). In this case we observe two counter-propagating structures over a background of Faraday waves.

Finally, Figure~\ref{fig:xt_osc}(d) depicts the response of the system to a larger forcing, namely $\epsilon = 1.32$. In this case six propagating structures are present at once in the system. Half of them are propagating clockwise, whereas the other half is moving counter-clockwise along the channel. All six share the same speed of $c = (342 \pm 16)$~{mm/s}. 

For forcing levels at or exceeding the threshold of $\epsilon~=~0.93$, propagating structures appear spontaneously and remain in the system for longer times. In prolonged experiments (8-minute duration, see Supplementary Material) the system spontaneously visits the various configurations available to the forcing level applied. 
Moreover, we observed that higher levels of forcing lead to similar behaviors albeit more robust and with longer-lasting propagating structures. These features collectively suggest the metastable nature of these states. In turn, the fact that these states seem to gain stability with increasing driving acceleration would indicate that the associated transition is subcritical.

\section{Discussion and conclusions}

Propagation in the form of drifting Faraday waves was initially reported in \cite{Douady1989} and discussed in \cite{martel2000dynamics}. It has been more recently observed in channel experiments with uniform forcing \cite{Guan2023} or with non-uniform forcing \cite{Urra2019, marin2023drifting}. The physical mechanism at the origin of this propagation lies in the existence of a large-scale or streaming flow that advects the structures. Larger structures were observed in a two-dimensional experiment

In our experiments, drifting waves are observed either before or in coexistence with the spontaneous propagation at higher velocity of one or more extended trains of localized superstructures in the form of traveling pulses on a periodic background in a one-dimensional Faraday wave. This is, to our knowledge, the first experimental work that has observed and measured the dynamics of these propagating localized structures. 

The dynamics of Faraday waves can be described through amplitude equations  \cite{miles1984parametrically, martel2000dynamics, Urra2019,marin2023drifting}. In particular, reference~\cite{martel2000dynamics} provides a comprehensive qualitative explanation of the observations reported in this letter. The authors study theoretically the problem of parametrically forced counterpropagating waves in a periodic annular domain. They consider the coupled amplitudes \( A \) and \( B \) of clockwise and counterclockwise traveling waves. Close to the instability onset, this coupling results in standing wave oscillations (\(|A| = |B|\)), consistent with Faraday waves. This formulation also accommodates instabilities that can break the reflection symmetry of the standing waves $(\theta \rightarrow -\theta)$ as forcing increases. \revision{Two distinct mechanisms arise: (a) one in which $|A| \neq |B|$ and they are uniform in space, \textit{i.e.}, they do not depend on $\theta$, and (b) another one in which the amplitudes $|A|$ and $|B|$ acquire spatial structure, \textit{i.e.}, $|A|= |A(\theta)|$ and $|B|= |B(\theta)|$. The first one (a) yields a state with a preferred drift direction. The second one (b) leads to spatially modulated structures. This latter type of pattern can be further classified into two categories: (i) $|A(\theta)|= |B(\theta)|$ corresponding to standing waves, and (ii) $|A(\theta)| \neq |B(\theta)|$ where both the Faraday waves and the modulation travel at different speeds. We believe that this second case corresponds to our experimental observations of propagating oscillons.}

\revision{Regarding propagation speed, the small drift of Faraday waves observed before oscillon appearance (Fig.~\ref{fig:xt_faraday}c) stems from the difference between the velocities of clockwise and counterclockwise travelling waves, each one of them moving at a speed of the order of the phase speed. This drift corresponds to case (a) of our previous classification. Since $|A| \ne |B|$ but $|A|\sim |B|$, the drift speed is slower than the group speed. Conversely, superstructures travel at a speed of the order of the group speed, as mentioned before. In this scenario ($|A(\theta)| \neq |B(\theta)|$), both drift types coexist in our experiments: the underlying Faraday pattern and localized structures travel at different speeds. The derivation in reference \cite{martel2000dynamics} incorporates an advection term, which suggests that the observed modulations can travel faster than the drifting Faraday waves --precisely what we observe experimentally. It is worth noting that their analysis applies sufficiently close to the threshold of the instability when the group velocity is comparable to the phase velocity, which corresponds to our experimental conditions. It is remarkable to note the similarity between the dynamics depicted in our Fig. \ref{fig:xt_osc} with Figs. 12 (bounded system) and 13 (periodic system) in the aforementioned reference.} 

Localized pulses over a one-dimensional pattern of Faraday waves have been previously observed experimentally only under spatially dependent forcing \cite{Urra2019, marin2023drifting}. The theoretical modeling of these cases has been developed considering an \textit{inhomogeneous} forcing in the parametrically driven nonlinear Schrödinger equation (PDNLS) for the envelope of the oscillation amplitude of a 1D parametric system. This theoretical model shows the existence of localized patterns whose characteristic extension is correlated to the spatial size of the forcing. It is also able to predict the drift of structures even when the forcing region is held fixed \cite{marin2023drifting}. When the forcing is homogeneous, numerical simulations \cite{Leon2015} have shown that the addition of an \textit{amending} nonlinear quadratic gradient term to the PDNLS leads to the spontaneous propagation of localized pulses over a patterned background. The modeling through the PDNLS equation differs from the formulation in terms of clockwise and counterclockwise traveling waves \cite{martel2000dynamics} since it takes only one amplitude equation that can be destabilized. The model in terms of two different amplitudes $A$ and $B$ allows the destabilization of $A$ and $B$ separately, providing a more extensive comprehension of the physical phenomena. 


In summary, our experiments reveal the emergence of propagating localized structures with a larger spatial extension, only attained previously using a spatially inhomogeneous forcing \cite{Urra2019,marin2023drifting}. Finally, in our prolonged experiments (see Suplementary Material) the system shows  complex dynamics over a much larger timescale, a behavior characteristic of similar structures studied in a much broader and fundamental context in physics. The detailed physical mechanism and theoretical description leading to the generation and dynamics of this type of localized structures remains elusive and warrants further investigation. Our findings offer valuable insights into the dynamics of propagating localized structures, which hold significance across diverse domains of physics including atmospheric and oceanic sciences, nonlinear optics, and quantum systems.

\section*{Acknowledgements}
The authors would like to thank S\'ebastien Gom\'e, Saliya Coulibaly, Marcel Clerc, and Juan Mar\'in for fruitful discussions, as well as Antonios Giannopoulos for his assistance. \revision{We are also grateful to the reviewers, whose thoughtful comments and suggestions substantially improved the quality of the discussion.} This work was partially supported by the IRP-IVMF between the CNRS and CONICET. S.K., and P.J.C. acknowledge support from grants PICT Nos.~2015-3530  and 2018-4298, and  UBACyT No.~20020170100508.


\bibliography{references.bib}
\bibliographystyle{unsrt}

\end{document}